\documentclass[final,3p,times]{elsarticle}

\usepackage{amssymb}
\usepackage[utf8]{inputenc}
\usepackage[T1]{fontenc}
\usepackage{amsmath}
\usepackage{amsthm}
\usepackage{tikz}
\usepackage{enumerate}

\usepackage{hyperref}
\hypersetup{
  colorlinks=true,
  linkcolor=blue,
  urlcolor=blue,
  citecolor=blue
}

\journal{Journal of Geometry and Physics}

\begin{document}

\begin{frontmatter}



\title{Superintegrable families of magnetic  monopoles with non-radial potential in curved background}

 \author[label1]{Antonella Marchesiello}
 \author[label2,label3]{Daniel Reyes}
 \author[label4]{Libor \v Snobl}
 \affiliation[label1]{organization={Faculty of Information Technology, Department of Applied Mathematics, Czech Technical University in Prague},
             addressline={Th\'akurova, 7},
             city={Prague 6},
             postcode={16000},
             country={Czech Republic}}

 \affiliation[label2]{organization={Departamento de F\'{\i}sica Te\'{o}rica, Facultad de Ciencias F\'{\i}sicas, Universidad Complutense de Madrid},
             city={Madrid},
             postcode={28040},
             country={Spain}}
   
 \affiliation[label3]{organization={Instituto de Ciencias Matem\'aticas},
             addressline={C/ Nicol\'as Cabrera, 13--15},
             city={Madrid},
             postcode={28049},
             country={Spain}}
             
\affiliation[label4]{organization={Faculty of Nuclear Sciences and Physical Engineering, Department of Physics, Czech Technical University in Prague},
             addressline={B\v rehov\'a, 7},
             city={Prague 1},
             postcode={11519},
             country={Czech Republic}}

\begin{abstract}
We review some known results on the superintegrability of monopole systems in the three-dimensional (3D) Euclidean space and  in the 3D generalized Taub-NUT spaces. We show that these results can be extended to certain curved backgrounds that, for suitable choice of the domain of the coordinates, can be related via conformal transformations to systems  in Taub-NUT spaces. These include the multi-fold Kepler systems as special cases. The curvature of the space is not constant and depends on a rational parameter that is also related to the order of the integrals. New results on minimal superintegrability when the electrostatic potential depends on both radial and angular variables are also presented.
\end{abstract}




\end{frontmatter}


\section{Introduction}
Superintegrable systems with magnetic monopole interaction have received much attention in mathematical physics.  In the three-dimensional (3D) Euclidean space,  any rotationally invariant electrostatic potential in the presence of a monopole  is at least minimally superintegrable \citep{Peres,MSW}, i.e. it possesses the minimal number of independent integrals that render it (Liouville) superintegrable.  Known maximally superintegrable exceptions (i.e. superintegrable systems possessing the maximal number of independent integrals allowed by their degrees of freedom) are both the MIC-oscillator \citep{McICis} and the hydrogen atom (MIC-Kepler monopole \citep{McICis,Zwan1968,IwaiTos}). Their superintegrability has been investigated and confirmed on various curved spaces, including e.g. 3D Taub-NUT spaces of Kepler-type and oscillator-type \citep{Iwai,HM2017} and the 3D sphere \citep{MonopoleSphere}.


Generalizations of superintegrable systems from flat to curved spaces have been extensively studied \citep{NP, Ballesteros, Tempesta2022}, resulting in the discovery of several families of superintegrable systems, often depending on a rational parameter related to the metric on the configuration space \citep{CR2020, Miller2010, Ranada2015, Ranada2021,LPW2012}. In the following, in analogy to the 3D generalization of the Tremblay-Turbiner-Winternitz (TTW) and Post-Winternitz (PW) systems obtained in \citep{Miller2010}, we present a family of superintegrable monopole systems in 3D curved background that include multi-fold Kepler systems \citep{Iwai2,Ballesteros} as special cases.

Let us consider the 4D generalized Taub-NUT metric \citep{Iwai}
\begin{equation}\label{TaubNUTmaetric}
ds^2= f(R)(dR^2 +  d \Theta^2 + R^2 \sin^2(\Theta)d\Phi^2)+ g(R)( d\Psi + \cos(\Theta) d \Phi)^2,
\end{equation}
for $R>0$, $\Theta\in(0,\pi)$, $\Phi\in[0,2\pi]$, $\Psi\in[0,4 \pi]$ and
\begin{equation}\label{Kepler-type}
f(R)=R^{\frac{1}{\nu}-2}(\alpha_1+ \beta_1 R^{\frac{1}{\nu}}), \qquad g(R)= \frac{(\alpha_1+ \beta_1 R^{\frac{1}{\nu}})R^{\frac{1}{\nu}}}{\alpha_2 R^{\frac{2}{\nu}}+ \beta_2 R^{\frac{1}{\nu}}+1},
\end{equation}
where $\nu, \alpha_i, \beta_i$ are real parameters. By the conservation of the quantity
$k=g(R)( \dot\Psi + \cos(\Theta) \dot\Phi)$, the corresponding free dynamics reduces to monopole systems in 3D Iwai-Katayama spaces, the so-called multi-fold Kepler systems introduced in \citep{Iwai2}. In particular, we have monopoles in oscillator-type and Kepler-type Taub-NUT spaces for $\nu=\frac12$ and $\nu=1$ \citep{Iwai}, respectively. For  $\nu=\frac12$, $\beta_1=0$, $\alpha_1=1$, we have  the MIC-oscillator monopole and for $\nu=1$, $\alpha_1=0$, $\beta_1=1$, the MIC-Kepler monopole system in Euclidean space.

Let us  now consider the 3D space  with the metric
\begin{equation}\label{metric}
ds^2=r^{-1}(\alpha_1+\beta_1 r)(dr^2 + r^2 m^2 d \theta^2 + r^2 \sin^2(\theta) d \varphi^2),  \;\; \theta\in(0,\pi),\; r>0,\; \varphi\in \left[0,\frac{2\pi}{\nu}\right]
\end{equation}
where $m$ is a nonvanishing constant. For $\nu=m^2=1$ we have just the Iwai-Katayama space for $\nu=1$. 
For $m^2\neq1$ we have a curved 3D space  with non-constant scalar curvature given by 
$\frac{2(1-m^{2})}{m^2 \left(\alpha_1+\beta_1 r\right) r }+\frac{3 \alpha_{1}^{2}}{2 \left(\alpha_1+\beta_1 r\right)^{3} r}.$
For $\alpha_1=0$ and $\beta_1=\nu=1$ this space is the one introduced in \citep{Miller2010}. However, the most natural choice $\nu=1$ for the domain of the angular variable $\varphi$ does not exhaust all the systems of physical interest.

The Cotton-York tensor of the metric \eqref{metric} vanishes identically for all $m$, therefore the underlying space is conformally flat.
  The canonical coordinate transformations
\begin{equation}\label{transf-conformally}
r= R^{m\delta}, \; \theta=\Theta, \; \varphi= m \delta \Phi,\; p_r= \frac{R^{1-m\delta} P_R}{m\delta},\; p_{\theta}=p_{\Theta},\; p_{\varphi}=\frac{P_{\Phi}}{m \delta},\;\;\delta\in\{-1,1\}
\end{equation}
brings the metric \eqref{metric} into the conformally flat form
\begin{equation}\label{metric-conformally}
ds^2=  m^2 R^{m\delta-2}(\alpha_1+\beta_1 R^{m\delta})(dR^2 + R^2 d \Theta^2 + R^2 \sin^2(\Theta) d \Phi^2), \; \; R>0,\; \Phi\in\left[0,\frac{2\pi}{m\nu}\right]
\end{equation}
that corresponds to  Iwai-Katayama spaces for $\nu=(m\delta)^{-1}$ (after proper renaming of $\alpha_1, \beta_1$). 

The corresponding multi-fold Kepler systems with magnetic monopole are known to be maximally superintegrable \citep{Ballesteros,Iwai2}. 
In the following section \ref{sec:system}, we introduce a Hamiltonian system with non-rotationally invariant potential in the metric space \eqref{metric} that in the special case when the non-radial part of the potential vanishes reduces to the family of multi-fold Kepler systems for $\nu=(m\delta)^{-1}$.  We will show in section \ref{sec:minimally} that for $m\in \mathbb Q$, this new family of systems has (at least) four independent integrals of motion and it is therefore minimally superintegrable. Thus, after breaking the radial symmetry of the multi-fold Kepler system, minimal superintegrability is preserved. The system exhibits three integrals of order at most quadratic in the momenta and the fourth integral is generally of higher order, with its order depending on $m$. We conclude this letter with some final remarks and conclusions in section \ref{sec:conclusions}.
 
\section{The systems}\label{sec:system}
We consider a Hamiltonian system in the 3D space with the metric \eqref{metric}
in the field of a magnetic monopole of strength $k$  
\begin{equation}\label{magfield}
B(r,\theta,\varphi)=k \sin(\theta) d\theta \wedge  d\varphi, \quad k\neq 0
\end{equation}
and an electrostatic potential of the form
\begin{equation}\label{potgen}
W(r,\theta, \varphi)=W_1(r) + \frac{W_2(\theta)}{r(\alpha_1+\beta_1 r)}.
\end{equation} 
Thus, its Hamiltonian function reads
\begin{equation}\label{Hamiltonian}
H=\frac{r}{2(\alpha_1+\beta_1 r)} \left((p_r^A)^2+\frac{(p_{\theta}^A)^2}{m^2 r^2}+ \frac{(p_{\varphi}^A)^2}{r^2 \sin ^2(\theta)}\right)+ W_1(r) + \frac{W_2(\theta)}{r(\alpha_1+\beta_1 r)}, \;\;\; m\in\mathbb R,\;m \neq0,
\end{equation}
where covariant expressions $p_j^A=p_j+ A_j$,  $j=r,\theta,\varphi$, for the momenta have been introduced. The spatial coordinates take values $r>0$, $\theta\in(0,\pi)$, $\varphi\in\left[0,\frac{2\pi}{\nu}\right]$, where $\nu>0$. The gauge for the magnetic field \eqref{magfield} can be chosen e.g. so that
\begin{equation}\label{gauge}
A_r=A_{\theta}=0, \;\; A_{\varphi}=(\ell-k \cos (\theta)),
\end{equation}
where we prefer to leave the constant $\ell\in\mathbb{R}$ arbitrary for the moment.

For $m^2=\nu=1$ and $\alpha_1=0$ we have a charged particle in the three-dimensional Euclidean space moving under the influence of the magnetic field \eqref{magfield} and potential \eqref{potgen}.

The system is rotationally invariant w.r.t to the angular variable $\varphi$ and indeed in proper gauge choice, e.g. \eqref{gauge}, the coordinate $\varphi$ is cyclic and $p_{\varphi}$ is a constant of motion. If we set $p_{\varphi}$ equal to a constant, the corresponding reduced system is separable in the coordinates $(r,\theta)$. This reflects itself in the existence of  the integrals 
\begin{equation}\label{integrals-gen}
X_1=p_\varphi^A+k \cos (\theta) \;\;\hbox{ and } \;\; X_2= (p_{\theta}^A)^2+m^2 \left(\frac{ (p_{\varphi}^A)^2}{\sin^2(\theta)}+ 2W_2(\theta)\right)
\end{equation}
that correspond to rotational invariance w.r.t to the angular variable $\varphi$ and conservation of a generalization of the total angular momentum which incorporates terms due to the presence of the magnetic field. 

Let us restrict our potential to the structure
\begin{equation}\label{W1superint}
W_1(r)= \frac{\alpha_2 r^2+\beta_2 r+k^2}{2 r (\alpha_1 +\beta_1  r)}
\;\;\hbox{ and }\;\;
W_2(\theta)=\frac{4 \left(a \cos ^2\left(\frac{\theta}{2}\right)+b \sin ^2\left(\frac{\theta}{2}\right)\right)+c}{ \sin ^2(\theta)},
\end{equation}
where $a,b,c,\alpha_i,\beta_i$ are constants. The motivation for this assumption comes from the fact that for $\nu=m^2=1$ we have a monopole system in Kepler-type Taub-NUT space and modified Hartmann potential \citep{Hartmann}.

Let us choose the gauge \eqref{gauge} with $\ell=k$ and use the fact that in this gauge choice $p_{\varphi}$ is a constant. Thus, we set  $p_{\varphi}=p_0$, $p_0$ constant and, after the canonical scaling
\begin{equation}\label{mscaling}
\theta\rightarrow \frac{\theta}{m}, \qquad p_{\theta}\rightarrow m p_{\theta},
\end{equation}
we see that  the system \eqref{Hamiltonian} reduces to the $2D$ system
\begin{equation}\label{2DHamiltonian}
\tilde H(p_r, p_{\theta}, r, \theta)=\frac{r}{2(\alpha_1+\beta_1 r)} \left(p_r^2+ \frac{p_{\theta}^2}{r^2}\right)+ \frac{1}{r(\alpha_1+\beta_1 r)}\left(\frac{\alpha}{\sin^2(\mu\theta)}+ \frac{\beta}{\cos^2(\mu\theta)}\right) + W_0(r), 
\end{equation}
where $m=\frac{1}{2\mu}$, $8\alpha=8 a+ 2c+ p_0^2$, $8\beta=8b+2c+(p_0+2 k)^2$, $W_0(r)= W_1(r)-\frac{k^2}{2r (\alpha_1 + \beta_1 r)}$. 
Thus, for $\alpha_1=0$, $\beta_1=1$,  the system becomes
the PW (Post-Winternitz) system \citep{PW2010}, known to be superintegrable for any $m\in\mathbb Q$. Therefore, besides $p_{\varphi}$, $X_2$ as in \eqref{integrals-gen} and the Hamiltonian, we have another integral coming from the 2D reduced system that renders the original system minimally superintegrable. The order of this integral depends on $m$.
To our knowledge, this relationship between the MIC-Kepler monopole and the  PW  system has not been noticed before, though the 3D system \eqref{Hamiltonian}-\eqref{W1superint} is known to be superintegrable for $c=0$, $m^2=\nu=1$ \citep{HM2017}. In the following, we will prove its minimal superintegrability for any value of the constants $a,b,c,\alpha_i,\beta_i$ and any $m\in\mathbb Q$.

By assuming $\nu=(m\delta)^{-1}$, $\delta\in\{-1,1\}$, the coordinate transformation \eqref{transf-conformally} brings the system \eqref{Hamiltonian} with potential \eqref{W1superint} to a generalized Taub-NUT  space  with new Hamiltonian  given by
\begin{equation}
 \mathcal H= \frac{R^{2-m\delta}}{2m^2(\alpha_1+\beta_1R^{m\delta})}\left((p_R^A)^2+\frac{(p_{\Theta}^A)^2}{R^{2}}+ \frac{(p_{\Phi}^A)^2}{R^2\sin^2(\Theta)}+ \frac{\alpha_2 R^{2m\delta}+\beta_2 R^{m\delta}+k^2}{R^2} + \frac{4 \left(a \cos ^2\left(\frac{\Theta}{2}\right)+b \sin ^2\left(\frac{\Theta}{2}\right)\right)+c}{R^2 \sin ^2(\Theta)} \right) \label{Hamiltonian-Taub}
\end{equation}
for $\theta\in(0,\pi)\;,\Phi\in\left[0,2\pi\right]$.
We will see in the following section \ref{sec:minimally} that
 system  \eqref{Hamiltonian}-\eqref{W1superint} is minimally superintegrable  for every rational value of $m$ and $\nu$. As a consequence, we derive also the superintegrability of system \eqref{Hamiltonian-Taub} and of \eqref{2DHamiltonian}. This extends the known results on the superintegrability of the PW  system and of the MIC-Kepler monopole system with modified Hartmann potential \eqref{W1superint} to the more general Taub-NUT metric space corresponding to \eqref{TaubNUTmaetric}-\eqref{Kepler-type}.
 
\section{Minimal superintegrability}\label{sec:minimally}
We will work from now on with the Hamiltonian \eqref{Hamiltonian} in the coordinates $(r,\theta,\varphi)$, with potential specified by \eqref{W1superint}.
As we saw in the previous section, the system \eqref{Hamiltonian} has integrals $X_1, X_2$ as in \eqref{integrals-gen}.
By following \citep{Miller2010}, we look for an additional integral that renders the system minimally superintegrable in the form
\begin{equation}\label{ansatz-integral}
X= M(r, H, X_2)-N(\theta, X_2,p_{\varphi})\equiv M(r)-N(\theta)
\end{equation}
where, in order to achieve $\{X, H\}=0$, $N$ and $M$ have to satisfy
\begin{equation}
\frac{p_{\theta}N'(\theta)-m^2 p_r r^2M'(r)}{m^2 r (\alpha_1 +\beta_1  r)}=0.\label{eqN} 
\end{equation}

By using the conservation of $p_{\varphi}=p_0$ and $H=E_0$, $X_2=E_1$, cf. \eqref{integrals-gen}, we can express $p_r$ and $p_{\theta}$ as $p_r(r,E_0,E_1)$ and $p_{\theta}(\theta, E_1, p_0)$, respectively. This allows us to solve equation \eqref{eqN} by separation of variables, looking for $N$ and $M$ such that
\begin{equation}
m^2 p_r(r,E_0,E_1) r^2 M'(r)=1
\;\;\hbox{ and }\;\;
 p_{\theta}(\theta, E_1, p_0)N'(\theta)=1.
 \end{equation}
Let us choose the gauge as in \eqref{gauge}. It is more convenient in the following computation to fix the constants so that $\ell=0$. This implies
$$M(r)=\pm \int \frac{1}{m r \sqrt{-\left(m^2 \left(r (\beta_2-2 E_0 (\alpha_1 +\beta_1  r)+\alpha_2 r)+k^2\right)\right)-E_1}}\, dr $$
and
$$N(\theta)=\pm \int \frac{\sin \theta}{\sqrt{ E_1-m^2 \left(4 a+4 b+2 c+p_0^2\right)-2 m^2 \cos(\theta) (2 a-2 b-k p_0)-\cos^2 (\theta)\left(E_1+k^2 m^2\right)} }\,d\theta \,.$$
By choosing the solution with a minus sign and integrating, we find
\begin{equation}
M(r)=-\frac{\hbox{arccos}\left(\mathcal T_1\right)}{m \sqrt{E_1+k^2 m^2}},\; \quad
N(\theta)=\frac{\arcsin\left(\mathcal T_2\right)}{\sqrt{E_1+k^2 m^2}},
\end{equation}
where 
\begin{eqnarray}
\mathcal T_1&=&\frac{m^2 \left(r (\beta_2-2 \alpha_1 E_0)+2 k^2\right)+2 E_1}{m r \sqrt{4 \alpha_1^2 E_0^2 m^2+8 \beta_1 E_0 E_1+4E_0 m^2 \left(2 \beta_1  k^2-\alpha_1 \beta_2\right)-4 \alpha_2 E_1+m^2 \left(\beta_2^2-4 \alpha_2 k^2\right)}},\\
\mathcal T_2&=&\frac{\cos (\theta) \left(E_1+k^2 m^2\right)-m^2 (-2 a+2 b+k p_0)}{\sqrt{2 m^4 \left(2 a^2-2 a (2 b+k (k+p_0))+2 b^2-2 b k (k-p_0)-c k^2\right)-E_1 m^2 \left(4 a+4 b+2 c-k^2+p_0^2\right)+E_1^2}}.
\end{eqnarray}
In order to obtain polynomial and globally defined integrals, we consider for $m=\frac{m_1}{m_2}\in\mathbb{Q}$, $g.c.d. (m_1,m_2)=1$,
$$\mathcal I= 2 i e^{\frac{1}{2} i \pi  m_2} \sin\left(m_1 \sqrt{E_1+k^2 m^2} \,  (  N(\theta)-M(r))\right),$$
that formally reads
$$\mathcal I=e^{i \pi m_2} \left(\sqrt{1-\mathcal T_2^2}+i \, \mathcal T_2\right)^{m_1} \left(\sqrt{1-\mathcal T_1^2}-i \, \mathcal T_1\right)^{m_2}-\left(\sqrt{1-\mathcal T_2^2}-i \, \mathcal T_2\right)^{m_1} \left(\sqrt{1-\mathcal T_1^2}+i \, \mathcal T_1\right)^{m_2}.$$
At this point, by using the fact that both square roots in the denominators of $\mathcal T_1$ and $\mathcal T_2$ are constants of motion, by multiplying $\mathcal I$ by suitable powers of those constants, we can obtain the simpler integral
\begin{eqnarray}
\mathcal X&=&\left(-2   \sqrt{ E_1+k^2 m^2} \, p_r+\frac{2 E_1+m^2 \left(r (\beta_2-2 \alpha_1  E_0)+2k^2\right)}{m r} i \right)^{m_2}\nonumber\\
&\cdot& \left( \sin (\theta) \sqrt{ E_1+k^2 m^2} \, p_{\theta} +\left(m^2 (2 a-2 b-k p_0)+ \cos (\theta) \left( E_1+k^2 m^2\right)\right)i \right)^{m_1}\nonumber\\
&-&\left(2  \sqrt{ E_1+k^2 m^2} \, p_r +\frac{ \left(m^2 \left(-2 \alpha_1  E_0 r+2 k^2+\beta_2 r\right)+2 E_1\right)}{m r} i \right)^{m_2}\nonumber\\
&\cdot& \left( \sin (\theta) \sqrt{ E_1+k^2 m^2} \, p_{\theta}- \left(m^2 (2 a-2 b-k p_0)+ \cos (\theta) \left( E_1+k^2 m^2\right)\right)i \right)^{m_1}, \label{PolX}
\end{eqnarray}
that except for the term $\sqrt{E_1+k^2 m^2}$ would be polynomial in the momenta, once substituting the explicit polynomial expressions of $E_0=H$ and $E_1=X_2$ in terms of the momenta. However, let us notice that given any complex numbers $z_1=a+ b i$ and $z_2=c+ d i$,
$$(-\overline{z_1})^{m_2} (\overline z_2)^{m_1}- z_1^{m_2} z_2^{m_1}=(-\overline{z_1})^{m_2} (\overline z_2)^{m_1}-(-1)^{m_2}(-z_1)^{m_2} z_2^{m_1}.$$
This expression equals to $-2i\hbox{Im}((-z_1)^{m_2} z_2^{m_1})$ for $m_2$ even and to 2Re$((-z_1)^{m_2} z_2^{m_1})$ for $m_2$ odd. By expanding the product $(-z_1)^{m_2} z_2^{m_1}$ we obtain
$$(-z_1)^{m_2} z_2^{m_1}=(-1)^{m_2}\sum_{j=0}^{m_1}\sum_{k=0}^{m_2}  i^{j+k}\left(\begin{array}{c}m_1\\j\end{array}\right)\left(\begin{array}{c}m_2\\k\end{array}\right) a^{m_2-k} b^k  c^{m_1-j} d^j.$$
Thus after expanding the products, the integral $\mathcal X$ would contain terms of the form $(E_1+k^2 m^2)^{\frac{m_1+m_2-(k+j)}{2}}$.  If $m_2$ is even, only exponents with $j+k$ odd would be present (i.e. we take only the imaginary part of the expansion as above). Since g.c.d.$(m_1,m_2)=1$ by assumption, for $m_2$ even $m_1$ must be odd. Therefore $m_1+m_2-(k+j)$ is even and only integer powers of $(E_1+k^2 m^2)$ appear in $\mathcal X$. Otherwise, for $m_2$ odd, only terms for $j+k$ even would be in the integral (i.e. only the real part).  Therefore, for $m_1$ odd, we have again that $m_1+m_2-(k+j)$ is an even number and only integer powers of $(E_1+k^2 m^2)$ are present in the integral. For $m_1$ even, $m_1+m_2-(k+j)$ is odd. However, we can divide $\mathcal X$ by $\sqrt{E_1+k^2 m^2}$ (which is a constant of motion) and obtain an integral with only even powers of  $(E_1+k^2 m^2)$. Therefore, we can always achieve a polynomial integral. 

\section{Final remarks and conclusions}\label{sec:conclusions}
In this manuscript, we proved the minimal superintegrability of the monopole system \eqref{Hamiltonian} with potential specified by \eqref{W1superint}. For particular values of its parameters, this system is related via conformal transformation to the MIC-Kepler, the MIC-oscillator and the more general multi-fold Kepler system \eqref{Hamiltonian-Taub} modified by the addition of a non-radial term in the potential. 

In recent years, there has been active research on (super)integrable systems in curved backgrounds or, alternatively, with position-dependent mass in Euclidean space \citep{Kalnins2011, Nikitin, Marquette2}, often resulting in the proof of  superintegrability of  Kepler-related systems and oscillator-related systems in curved spaces \citep{Ranada2021,HM2017, Miller2010, Ballesteros2}. Here we saw that this is also true in presence of a magnetic monopole in the metric space corresponding to \eqref{metric}. This extends known results for Kepler-type and monopole-type systems in generalized Taub-NUT spaces and shows that minimal superintegrability is preserved even if the non-radial term specified by \eqref{W1superint} is added to the potential. 

We attempted to search for a fifth integral by the method used in the previous section, however with the negative result that such an integral would be polynomial only for $a=b=c=0$, i.e. for vanishing angular term in the potential, and globally defined only for $\nu\in\mathbb Q$. Therefore we just obtain the known result on the maximal superintegrability of the multi-fold Kepler system. Numerical investigation seems to confirm that for generic values of the parameters $a,b,c$ there can exist bounded orbits which do not close, therefore we expect the system  to be in general only minimally superintegrable.

Let us notice that given the Hamiltonian in the form
\begin{equation}
H=\frac12 \, f(r) \, \left(p_r^2+ \frac{ L}{r^2}\right),
\end{equation}
where $L$ is a constant of motion, by scaling 
$L \rightarrow \frac{L}{m^2}$
 the new Hamiltonian
\begin{equation}
\tilde H=\frac12 \, f(r) \, \left(p_r^2+ \frac{ L}{m^2 r^2}\right)
\end{equation}
is still integrable \citep{LPW2012}. However, it corresponds to a system in general with different metric and there is no guarantee that superintegrability is preserved. Therefore, the integrability of the system \eqref{Hamiltonian} with potential \eqref{W1superint} for any nonvanishing $m$ comes from the integrability of the corresponding  Kepler-type monopole system for $m=1$. However, the superintegrability of the system for general value of $m$ can not be deduced as a consequence of the superintegrability of the original system, as to our knowledge there is no proof that all the integrals survive the scaling, though this is often the case if $m$ is rational for many examples known \citep{LPW2012}. 
It might be worth further investigation in this direction.

Let us conclude by mentioning that the system \eqref{Hamiltonian-Taub} is related to an interesting family of systems conjectured to be superintegrable \citep{Tempesta2022}. In the limit for vanishing strength of the monopole and for $\alpha_1=c=0$, the Hamiltonian \eqref{Hamiltonian-Taub}  is included in it for special choices of the parameters therein. 
In this perspective, it might come as a surprise that only the PW system appeared here, as a 2D reduction of the system \eqref{Hamiltonian}, though also the TTW system is included in the family of systems presented in \citep{Tempesta2022}. The TTW system can indeed be obtained from Hamiltonian \eqref{Hamiltonian-Taub} (canonically conjugated to \eqref{Hamiltonian}) after setting $m=\frac{2}{\delta}$, $p_{\varphi}$ to constant and then performing the scaling \eqref{mscaling}.  This seems to suggest that also in presence of a magnetic field, a similar family of superintegrable systems as the one introduced in \citep{Tempesta2022} that includes the monopole might exist.

\section*{Acknowledgements}
The authors thank C.M. Chanu and P. Tempesta for fruitful discussions.
AM and L\v{S} received support from the Czech Ministry of Education, Youth and Sports RVO 68407700. DR acknowledges the financial support of EXINA S.L. and of the UCM grants for short stays of predoctoral contract holders. The research was initiated during his ERASMUS+ stay at CTU, Prague, to which he thanks for the warm hospitality.








\end{document}